\documentclass{article}

\usepackage{url,a4wide}
\usepackage{amsmath,amsfonts,amssymb}
\usepackage{stmaryrd}
\usepackage{xspace}

\newcommand{\lbrakk}{\llbracket}
\newcommand{\rbrakk}{\rrbracket}
\newcommand{\noteq}{\neq}

\newcommand{\SpecRel}{\textsc{SpecRel}\xspace}
\newcommand{\GenRel}{\textsc{GenRel}\xspace}
\newcommand{\AccRel}{\textsc{AccRel}\xspace}

\newcommand{\Body}{\texttt{Body}\xspace}
\newcommand{\SpaceTime}{\texttt{SpaceTime}\xspace}
\newcommand{\Quantities}{\texttt{Quantities}\xspace}
\newcommand{\WorldView}{\texttt{WorldView}\xspace}
\newcommand{\Points}{\texttt{Points}\xspace}
\newcommand{\Vectors}{\texttt{Vectors}\xspace}
\newcommand{\Lines}{\texttt{Lines}\xspace}
\newcommand{\Planes}{\texttt{Planes}\xspace}
\newcommand{\Cones}{\texttt{Cones}\xspace}

\newcommand{\Axiom}[1]{\textbf{Ax{#1}}}
\newcommand{\Code}[1]{\texttt{#1}\xspace}

\title{Using Isabelle to verify special relativity,\\with application to hypercomputation theory}
\author{
  MIKE STANNETT
  \\Computer Science Department, University of Sheffield 
  \\Regent Court, 211 Portobello, Sheffield S1 4DP, United Kingdom
  \\\url{m.stannett@dcs.shef.ac.uk}
\\[6pt]
  ISTV\'AN N\'EMETI
  \\Alfr\'ed R\'enyi Institute of Mathematics of the Hungarian Academy of Sciences
  \\P.O. Box 127, Budapest, 1364, Hungary
  \\\url{nemeti.istvan@renyi.mta.hu}
}

\begin{document}

\maketitle

\begin{abstract}
Logicians at the R{\'e}nyi Mathematical Institute in Budapest have spent several years developing versions of relativity theory (special, general, and other variants) based wholly on first order logic, and have argued in favour of the physical decidability, via exploitation of cosmological phenomena, of formally undecidable questions such as the Halting Problem and the consistency of set theory.

The Hungarian theories are very extensive, and their associated proofs are intuitively very satisfying, but this brings its own risks since intuition can sometimes be misleading. As part of a joint project, researchers at Sheffield have recently started generating rigorous machine-verified versions of the Hungarian proofs, so as to demonstrate the soundness of their work. In this paper, we explain the background to the project and demonstrate an Isabelle proof of the theorem ``No inertial observer can travel faster than light''. 

This approach to physical theories and physical computability has several pay-offs: (a) we can be certain our intuition hasn't led us astray (or if it has, we can identify where this has happened); (b) we can identify which axioms are specifically required in the proof of each theorem and to what extent those axioms can be weakened (the fewer assumptions we make up-front, the stronger the results); and (c) we can identify whether new formal proof techniques and tactics are needed when tackling physical as opposed to mathematical theories.
\end{abstract}

\noindent {Categories and Subject Descriptors:} 
{F.4.1}
         [\textbf{Mathematical Logic and Formal Languages}]
         {Mathematical Logic}---\textit{Mechanical theorem proving};
{J.2}
         [\textbf{Computer Applications}]
         {Physical Sciences and Engineering}---\textit{Physics}
\\[6pt]
\noindent General Terms: {Theory, Verification}
\\[6pt]
\noindent Additional Key Words and Phrases: {First-order relativity theory, hypercomputation, physics and computation}

\section{Introduction}
\label{sec:introduction}

In his seminal analysis of computation, Turing  \cite{Tur36} discussed the nature of human computation, and showed that certain tasks -- most famously, the Halting Problem (HP) -- are not decidable by computational means. Subsequent theoretical investigation by various researchers suggests, however, that physical systems may exist which can in fact decide HP by exploiting cosmological phenomena \cite{Hog92,EN93,EN02,Hog04,Man10,ANS12}. This claim is, of course, highly controversial; we therefore begin by explaining the loophole in Turing's analysis which allows `hypercomputational' systems of this kind to be designed \cite{Sta06,Sta13}. 

We then focus on one particular scheme for cosmological hypercomputation \cite{EN02}, and consider the extent to which it rests on secure logical foundations. Doing so will require us to explain recent work by the Hungarian team of Andr\'eka et al, who have formalised a series of relativity theories (including special and general relativity) using first-order logic \cite{AMN04,AMNS08}. These first-order foundations ensure that their theories are easy to reason about, but also introduce a number of nonstandard features. We have, therefore, recently started a joint project verifying their theories using the Isabelle proof assistant \cite{Isabelle}. We explain our approach below, and outline an Isabelle proof of the well-known statement ``No inertial observer can travel faster than light'' \cite{Ein20,AMNS12}. Finally, we summarise the work that remains to be done, and invite participation in the solution of several open questions.

\section{Circumventing Turing's analysis}
\label{sec:circumventing-turings-analysis}

Turing's \cite{Tur36} analysis of (human) computation provides a convincing demonstration that certain problems cannot be solved by computational means. In particular, if $P_0, P_1, P_2, \dots$ is a fixed enumeration of all programs\footnote{For simplicity, we will think of programs as being written in a modern high-level language, running on a standard PC with access to unbounded memory.} that take a single natural number as input, it is not possible to compute the function $\mathit{HP} \colon \mathbb{N} \times \mathbb{N} \to \{ \mathit{yes}, \mathit{no} \}$ given by
\[
  \mathit{HP}(m,n) = \begin{cases}
    \mathit{yes} & \text{ if $P_m(n)$ will eventually halt } \\
    \mathit{no}  & \text{ otherwise }
  \end{cases}
\]

Powerful as it is, Turing's analysis is nonetheless susceptible to attack due to an unexamined assumption built into his description of human computation. For, as he explains \cite{Tur50}:
\begin{quote}
The human computer is supposed to be following fixed rules; he has no authority to deviate from them in any detail. We may suppose that these rules are supplied in a book, which is altered whenever he is put on to a new job. He has also an unlimited supply of paper on which he does his calculations. He may also do his multiplications and additions on a ``desk machine,'' but this is not important.
\end{quote}
\noindent In fact, the consequences of using a ``desk machine'' cannot be so readily dismissed, because this implies that the computation may involve coordination between two physically separated agents (the human and the machine) \cite{Sta13}. Being physically separated, the two agents may be subject to different forces and accelerations, and this can affect the rate at which they perceive each other's clocks to be running. This in turn provides scope for extreme computational speed-up, to the extent that HP becomes solvable. For example, astronomical observations suggest the presence of a massive slowly rotating (``\textit{slow Kerr}'') black hole at the centre of the Milky Way \cite{GET+09}. Such black holes are associated, in relativity theory, with a computationally useful spacetime geometry (\textit{Malament-Hogarth spacetime} \cite{EN93}), containing a worldline $w$ and a point $p$ (not on $w$), with the following properties:
\begin{itemize}
\item
	$w$ has infinite proper length;
\item
  it is possible to send a signal to $p$ from any point along $w$.
\end{itemize}

Suppose, then, that we are given $m$ and $n$, and want to determine whether or not P $\equiv P_m(n)$ will eventually halt. We send a PC along $w$ having first loaded an interpreter with behaviour:

\begin{verbatim}
  run P;
  send a signal to spacetime location p
\end{verbatim}

\noindent If P doesn't halt, the second instruction will never be reached, and no signal will be sent. On the other hand, because $w$ has infinite proper length, the PC has unbounded time available to it for its computation, and so P has enough time to run to completion if this is its underlying behaviour. Consequently, a signal will arrive at $p$ if and only if $P_m(n)$ eventually halts. It is therefore enough for us to follow a trajectory that takes us through $p$. When we arrive there, we look for the presence of the signal, saying $\mathit{yes}$ if the signal is present, and $\mathit{no}$ otherwise.

\section{Logical foundations}
\label{sec:logical-foundations}

We now turn to Andr\'eka et al's  \cite{AMN04,AMNS12} first-order formalisation of relativity theory. This focus on first-order logic (FOL) is motivated by several important considerations. Foremost is the Hungarian team's desire to demystify relativity theory by expressing its postulates and conclusions in a form that is intelligible to as large an audience as possible. By choosing simple language and a very simple axiom system, the underlying assumptions of the theory are made as straightforward as possible (see Sect. \ref{sec:axioms}), while the use of first-order logic and its simple reliance on Modus Ponens makes it relatively easy for newcomers to follow the proofs. Having reformulated relativity in purely logical terms, the group is also able to investigate which axioms underpin which results and which are superfluous. Given the physical nature of the theory in question, this information can then be reflected back into physics: if an axiom plays no role in establishing an experimentally observed result, then that result can neither support nor undermine the validity of the axiomatic property in question.

It is important to note, however, that the use of first-order logic has important consequences when attempting to model physical phenomena, because FOL is not powerful enough to characterise the real number field, $\mathbb{R}$ -- the numbers typically used to represent coordinates, masses, and so forth, in physical models. Consequently, many of the real-number properties we take for granted in physics, like the existence of limits of bounded sequences, are unavailable in a rigorous first-order logical proof.\footnote{For completeness, we note that this difficulty can be solved within FOL by focussing attention on definable sets.} For example, the statement that any decreasing sequence of real numbers, bounded below, has a greatest lower bound is not a first order statement, because it refers to ordered \emph{sets} of real values.\footnote{There are fields which have the same first-order properties as $\mathbb{R}$, but which contain infinitesimals. In such a field, the bounded decreasing sequence $\frac{1}{1} > \frac{1}{2} > \frac{1}{3} > \dots$ has no greatest lower bound. For suppose $\alpha$ were its greatest lower bound; then given any positive infinitesimal $\epsilon$, the value $(1+\epsilon)\alpha$ would be a slightly larger lower bound, thereby contradicting the definition of $\alpha$.} Moreover, as Andr\'eka and her colleagues have shown, many interesting theorems can be proven using less restrictive fields like the rationals, $\mathbb{Q}$, for which the real-number property \emph{every positive number has a positive square root} fails\footnote{The statement cited is first-order: $(\forall x).((x > 0) \to (\exists y . ((y > 0) \land (y \times y = x))))$.} (such fields are said to be \emph{non-Euclidean}), cf. \cite{Sze09}.

\subsection{The need for formal verification}
\label{sec:need-for-verification}

Given that ``first-order numbers'' need not exhibit the properties typically expected of them by physicists, it is important that we treat traditional explanations of relativistic phenomena with caution. To this end, and as part of a Royal Society International Exchanges Scheme project, researchers in Sheffield joined forces with the Hungarian team at the start of 2012, to develop a comprehensive formal framework for relativity theory, with full machine-verification of all derived theorems. To the best of our knowledge, this is the first time such a large-scale physical theory has been treated in this way (but cf. \cite{GS11,BS12}), and it is hoped that the lessons learned will be useful in extending the approach more widely. The project has been planned in four main stages, and it is hoped that the end result will be a formal machine-verified proof of the controversial claim that the power of a computational system depends on the nature of its spacetime environment, with super-Turing capabilities emerging in the context of more complex spacetime geometries.

The project itself has four broad aims:
\begin{enumerate}
\item
	Implement first-order axiomatizations of general relativity using the proof assistant Isabelle \cite{Isabelle};
\item
	Add a general model of computational mobility to the theory, to enable the modelling of computations carried out by machines travelling along specific spacetime trajectories;
\item
	Consider how the power of these computational systems changes according to the underlying topology of spacetime \cite{UCNC12};
\item
	Select a recursively uncomputable problem \textbf{P} (for example, the Halting Problem) and machine-verify the following claims:
	\begin{enumerate}
	\item
		in simpler relativistic settings, \textbf{P} remains uncomputable;
	\item
		in some spacetimes, \textbf{P} can be solved.
	\end{enumerate}
\end{enumerate}

Taken together, these steps are intended to add weight to the claim that the computational power of a device depends on the physical setting in which it finds itself.

\section{The theories and their implementation}
\label{sec:theories}

There are various versions of relativity theory, depending on what is being modelled. For special relativity (\SpecRel) the two key axioms (suitably formalised) are \cite{Ein20}:

\begin{quote}
\textbf{Principle of relativity:}
The laws of nature are the same for every inertial observer;
\end{quote}

\begin{quote}
\textbf{Light postulate:}
Any ray of light moves in the `stationary' system of coordinates with the determined velocity $c$, whether the ray be emitted by a stationary or by a moving body;
\end{quote}

\noindent while for general relativity (\GenRel) we add the

\begin{quote}
\textbf{Equivalence Principle:}
It is not possible to distinguish between the effects of acceleration and those of gravity.
\end{quote}

In addition to special and general relativity, Sz\'ekely and his colleagues have made a detailed study of \emph{accelerated observers} (with or without the equivalence principle in place). The corresponding theory, \AccRel, provides a convenient stepping stone from special to general relativity \cite{Sze09}.

Our Isabelle implementation\footnote{The files referred to in this paper are available from \url{http://www.dcs.shef.ac.uk/~mps/isabelle/noFTLobserver}.} has been constructed in three parts, a program structure that ensures that different versions of relativity theory can easily be added later. For example, to add \GenRel we would simply add a new file \texttt{GenRel.thy} which merges the required axiom classes and includes proofs of relevant theorems. We focus here on the first-order theory \SpecRel of special relativity. This theory is 2-sorted, the sorts being \Quantities (the values used to specify coordinates, speeds, masses, etc) and \Body (\emph{bodies} or \emph{test particles}).

\subsection{Background geometry (\texttt{SpaceTime.thy}, approx. 830 lines)}
\label{ref:background-geometry}
This Isabelle/HOL code file models the geometric structures common to all models of spacetime (\Vectors, \Points, \Lines, \Planes, \Cones), each represented as a separate record structure with axioms attached. The axioms describe basic geometric relationships including, for example, what it means for three points to be collinear, what it means for two vectors to be orthogonal, and so forth. In particular, a key lemma for our main proof is the assertion that distinct parallel lines cannot meet (the proof is by contradiction). Having defined these classes, we take \SpaceTime to be their conjunction:

\begin{verbatim}
class SpaceTime = Quantities + Vectors + Points + Lines + Planes + Cones
\end{verbatim}

The set of \Quantities is assumed to carry an ordered field structure. We shall sometimes need to assume that the field is also Euclidean -- i.e., that square roots exist for positive values -- but this is not a general requirement, so it will be added as a separate axiom class later. Since Isabelle/HOL already includes a suitable class, the implementation of \Quantities is particularly simple:

\begin{verbatim}
class Quantities = linordered_field
\end{verbatim}

For simplicity we assume that spacetime is $(1+3)$-dimensional (one time dimension + three space dimensions), so that \Points and \Vectors are both specified as 4-tuples of \Quantities. In more complex relativity theories, we allow both the number of space dimensions, and the number of time dimensions, to vary. Lines are specified by giving a point (the line's \emph{basepoint}) and a vector (its \emph{direction}), while planes are specified by a basepoint and two vectors. 

Because we are dealing here with special relativity, all lightcones can be considered to be `upright' (for general relativity we need to allow cones that are `tilted' by curvature effects); each cone can therefore be specified by giving a point (its \emph{vertex}) and a quantity (its \emph{slope}). However, the freedom with which we can specify quantities has certain concomitant side-effects, and these need to be taken into account. In real-number physics, we would consider the slope of the cone
\[
  x^2 + y^2 + z^2 = \alpha t^2 \qquad \text{ where $\alpha > 0$ }
\]
\noindent to be $\sqrt{\alpha}$, but when \Quantities is non-Euclidean we cannot be certain that $\sqrt{\alpha}$ is defined. Consequently, we take the slope of the cone to be $\alpha$ rather than $\sqrt{\alpha}$, and adjust all associated formulae and proofs accordingly.

\subsection{Axioms (\texttt{Axioms.thy}, approx. 260 lines)}
\label{sec:axioms}
This file includes various axioms used by the Hungarian group, each implemented as a separate class. Different relativity theories can then be constructed by merging the relevant axiom classes and omitting those that are not required; we focus here on the axioms that will be needed to specify \SpecRel. 

The axioms describe the events in which bodies can participate, and how their descriptions change from one observer's viewpoint to another. Here, a \Body can be either a \emph{photon} (which always travels at constant speed) or an \emph{inertial observer} (which always travels at constant speed, and in addition is capable of making observations). Since we do \emph{not} assume a priori that the classes of photons and inertial observers are disjoint, we represent bodies using an Isabelle/HOL record structure:

\begin{verbatim}
record Body =
  Ph :: "bool"
  IOb :: "bool"
\end{verbatim}

For more complex relativistic theories we also need to consider non-inertial observers (those which can accelerate), as well as more general types of body, and in this regard the use of Isabelle/HOL record structures is particularly convenient, since we can easily extend the \Body record structure to include new descriptions. The distinction between inertial observers and more general body types emerges in these more advanced theories. For example, we demonstrate below that inertial observers can never travel faster than (what they consider to be) the speed of light, but this property need \emph{not} be provable of more general bodies \cite{NS12,Sze12}.

In addition to the ordered field axioms associated with \Quantities, \Code{SpecRel} is formally generated using just the four axioms described below (\Code{AxPh}, \Code{AxEv}, \Code{AxSelf}, \Code{AxSym}), but in practice we have found it sensible to replace \Quantities with a larger \WorldView class (below) so as to have available the necessary abbreviations and functions. This simplifies proofs considerably. Moreover, our proof that inertial observers cannot travel faster than light requires us to find the intersection of a line with a cone, and this in turn requires the existence of square roots -- we have therefore included the Euclidean axiom (\Code{AxEuclidean}). Finally, we make use of various additional properties of cones, lines and planes (given in \Code{SpaceTime.thy}). These define various relatively complicated concepts, such as what it means for a plane to be tangent to a (light)cone:

\Code{
~
\\class Cones = Quantities + Lines + Planes +
\\fixes 
\\  \indent tangentPlane :: "'a Point $\Rightarrow$ 'a Cone $\Rightarrow$ 'a Plane"
\\\\assumes (* The basepoint of the tangent-plane-at-e is e *)
\\  \indent AxTangentBase: "pbasepoint (tangentPlane e cone) = e"
\\\\and (* The tangent plane contains the vertex *)
\\  \indent AxTangentVertex: "inPlane (vertex cone) (tangentPlane e cone)"
\\\\and (* The tangent plane meets the cone in a line *)
\\  \indent AxConeTangent: "(onCone e cone) $\longrightarrow$
\\  \indent \indent (inPlane pt (tangentPlane e cone) $\land$ onCone pt cone) 
\\  \indent \indent \indent $\longleftrightarrow$ collinear (vertex cone) e pt)"
\\\\and (* The tangent plane is tangential to all cones with vertex 
\\  \indent \indent in that plane, and the intersection lines are parallel. *)
\\  \indent AxParallelCones: "(onCone e econe $\land$ e $\noteq$ vertex econe 
\\  \indent \indent $\land$ onCone f fcone $\land$ f $\noteq$ vertex fcone
\\  \indent \indent $\land$ inPlane f (tangentPlane e econe))
\\  \indent \indent $\longrightarrow$  (samePlane (tangentPlane e econe) (tangentPlane f fcone)
\\  \indent \indent $\land$ ((lineJoining (vertex econe) e) $\parallel$ (lineJoining (vertex fcone) f)))"
\\\\and (* If f is outside a cone, there is a tangent plane to that cone which 
\\  \indent \indent contains f. The tangent plane is determined by some e lying on 
\\  \indent \indent the intersection line with the cone. *)
\\  \indent AxParallelConesE: "outsideCone f cone $\longrightarrow$ ($\exists$e.(onCone e cone 
\\  \indent \indent $\land$ e $\noteq$ vertex cone $\land$ inPlane f (tangentPlane e cone)))"
\\
}

\subsection*{\Axiom{Euclidean}}
This axiom states that every positive quantity has a positive square root, and defines the \Code{sqrt} function.

\Code{
~
\\class AxEuclidean = Quantities +
\\assumes
\\  \indent AxEuclidean: "(x $\geq$ (0::'a)) $\Longrightarrow$ ($\exists$r. ((r $\geq$ 0) $\land$ (r*r = x)))"
\\begin
\\  \indent fun sqrt :: "'a $\Rightarrow$ 'a" where
\\  \indent\indent "sqrt x = (SOME r. ((r $\geq$ (0::'a)) $\land$(r*r = x)))"
\\end
\\
}

Notice, however, that we do \emph{not} assume that the positive square root is uniquely defined (instead, this is a theorem). Consequently, even though \Code{sqrt} is defined using the \textbf{fun} keyword, it is not in fact defined to be a function, because the use of \Code{SOME} technically allows a different value to be selected each time \Code{sqrt} is referenced.

\subsection*{The WorldView relation}
Two key features of first-order relativity theory are the \emph{worldview relation} (\Code{W}) and the \emph{worldview transformation} (\Code{wvt}).

\Code{
~
\\class WorldView = SpaceTime +
\\fixes
\\  \indent (* Worldview relation *)
\\  \indent W :: "Body $\Rightarrow$ Body $\Rightarrow$ 'a Point $\Rightarrow$ bool" ("\_ sees \_ at \_")
\\and
\\  \indent (* Worldview transformation *)
\\  \indent wvt :: "Body $\Rightarrow$ Body $\Rightarrow$ 'a Point $\Rightarrow$ 'a Point"
\\assumes
\\  \indent AxWVT: "$\lbrakk$ IOb m; IOb k $\rbrakk \Longrightarrow$ (W k b x $\longleftrightarrow$ W m b (wvt m k x))"
\\and
\\  \indent AxWVTSym: "$\lbrakk$ IOb m; IOb k $\rbrakk \Longrightarrow$ (y = wvt k m x  $\longleftrightarrow$  x = wvt m k y)"
\\begin
\\end
\\
}

The relation \Code{W} tells us which bodies an inertial observer $m$ sees at each spacetime location. Thus, \Code{W m b p} is \Code{True} precisely when \Code{m} considers the body (whether inertial observer or photon) \Code{b} to be present at location \Code{p}. We can use \Code{W} to define various standard concepts; for example, the worldline of \Code{b} (from \Code{m}'s point of view) is simply the set \Code{\{p . W m b p\}}.

The worldview transformation tells us how one observer's viewpoint is related to another. As \Code{AxWVT} explains, if \Code{wvt m k x} is \Code{y}, this means that whatever \Code{k} sees at \Code{x}, \Code{m} sees at \Code{y}.

\subsection*{\Axiom{Ph}}
The \emph{photon axiom} says that for any inertial observer, the speed of light ($c$) is the same in every (spatial) direction everywhere and is positive. Furthermore, it is possible to send out a light signal in any (spatial) direction. (The auxiliary functions \Code{space2} and \Code{time2} give the squared spatial and temporal separations, respectively, of two spacetime locations \Code{x} and \Code{y}.)

\Code{
~
\\class AxPh = WorldView +
\\assumes
\\  AxPh: "IOb(m) 
\\  \indent $\Longrightarrow$ ($\exists$v. ( (v $>$ (0::'a)) $\land$ ( $\forall$x y . ( 
\\  \indent \indent ($\exists$p. (Ph p $\land$ W m p x $\land$ W m p y)) 
\\  \indent \indent \indent $\longleftrightarrow$ (space2 x y = (v * v)*(time2 x y)) 
\\  \indent ))))"
\\begin
\\  \indent fun c :: "Body $\Rightarrow$ 'a" where
\\  \indent \indent "c m = (SOME v. ( (v $>$ (0::'a)) $\land$ ( $\forall$x y . ( 
\\  \indent \indent \indent $\exists$p. (Ph p $\land$ W m p x $\land$ W m p y)) 
\\  \indent \indent \indent \indent $\longleftrightarrow$ (space2 x y = (v * v)*(time2 x y)) 
\\  \indent \indent )))"
\\
\\  \indent fun lightcone :: "Body $\Rightarrow$ 'a Point $\Rightarrow$ 'a Cone" where
\\  \indent \indent lightcone m v = mkCone v (c m)"
\\
\\  \indent (* various lemmas follow that are not included here *)
\\
}

Notice, however, that the speed of light is not assumed to be the same for all observers: the value $c$ is therefore parametrised according to the inertial observer in question. As before, the use of \Code{SOME} suggests that \Code{c m} need not be uniquely defined, but uniqueness becomes provable within \Code{SpecRel} due to the inclusion of additional axioms. Note also that \Code{c p} is technically specified when \Code{p} is a photon; but in this case the precondition required to establish the value's existence cannot be established using \Code{AxPh}. In this way we avoid the (non)question ``at what speed does one photon consider another photon to be travelling?''

\subsection*{\Axiom{Ev}}
The \emph{event axiom} says that all inertial observers are participating in the same universe -- if one observer sees two bodies meeting at some spacetime location, they all see them meeting (though they may disagree as to where that meeting takes place).

\Code{
~
\\class AxEv = WorldView +
\\assumes
\\  \indent AxEv: "$\lbrakk$ IOb m; IOb k $\rbrakk$ $\Longrightarrow$  ($\exists$y. ($\forall$b. (W m b x  $\longleftrightarrow$ W k b y)))"
\\begin
\\end
\\
}

\subsection*{\Axiom{Self}}

The \emph{self axiom} says that inertial observers consider themselves to be stationary in space (so they consider their worldline to be the time axis)

\Code{
~
\\class AxSelf = WorldView +
\\assumes
\\  \indent AxSelf: "IOb m  $\Longrightarrow$  (W m m x) $\longrightarrow$ (onAxisT x)"
\\begin
\\end
\\
}

\subsection*{\Axiom{Sym}}

The \emph{symmetry axiom} says that inertial observers agree as to the spatial distance between two spacetime events if these two events are simultaneous for both of them.

\Code{
~
\\class AxSym = WorldView +
\\assumes
\\  \indent AxSym: "$\lbrakk$ IOb m; IOb k $\rbrakk$ $\Longrightarrow$
\\  \indent \indent (W m e x $\land$ W m f y $\land$ W k e x' $\land$ W k f y' $\land$
\\  \indent \indent \indent tval x = tval y $\land$ tval x' = tval y' )
\\  \indent \indent $\longrightarrow$ (space2 x y = space2 x' y')"
\\begin
\\end
\\
}

\subsection{\SpecRel (\texttt{SpecRel.thy}, approx. 340 lines)}
\label{sec:specrel}

This file defines the theory \SpecRel, 

\begin{verbatim}
class SpecRel = WorldView + AxPh + AxEv + AxSelf + AxSym
(*
  The following proof assumes that the quantity field is Euclidean.
*)
  + AxEuclidean
(*
  We also assume for now that lines, planes and lightcones are
  preserved by the worldview transformation. This can be proven.
*)
  + AxLines + AxPlanes + AxCones
\end{verbatim}

\noindent together with our proof of the standard claim that no inertial observer can travel faster than the speed of light.

\section{The proof}
\label{sec:the-proof}

The statement we wish to prove (``no inertial observer can travel faster than light'') can be formalised as:

\Code{
~
\\lemma noFTLObserver:
\\  \indent assumes iobm:  "IOb m"
\\  \indent and \quad iobk:  "IOb k"
\\  \indent and \quad mke:   "m sees k at e"
\\  \indent and \quad mkf:   "m sees k at f"
\\  \indent and \quad enotf: "e $\noteq$ f"
\\shows \quad "space2 e f $\leq$ (c m * c m) * time2 e f"
\\
}

To see why, notice that the statement ``\Code{k} cannot travel faster than light'' is meaningless as it stands. We need to say \emph{in whose opinion} this statement is true, since the speed of light might depend on the observer. We therefore have to introduce a second inertial observer, \Code{m}, in whose opinion the judgment is to be made. To find the speed at which \Code{k} is moving, \Code{m} needs to observe \Code{k} at two different locations, \Code{e} and \Code{f}, and then determine the (square of the) ratio of the associated spatial and temporal separations.

The proof itself is in five basic stages.

\subsection*{Step 1. Assume the converse}
Suppose \Code{k} is going faster than light (FTL) from \Code{m}'s viewpoint:

\Code{
~
\\
\\  \indent assume converse: "space2 e f > (c m * c m) * time2 e f" 
\\
}

Informally, we are saying that \Code{f} lies outside \Code{m}'s lightcone at \Code{e}.

\subsection*{Step 2. Consider the cone at \Code{e}}

Consider \Code{m}'s lightcone at \Code{e}, and note that \Code{e} is itself on this cone (since it is the cone's vertex).

\Code{
~
\\  \indent def eCone $\equiv$ "mkCone e (c m)"
\\  \indent have e\_on\_econe: "onCone e eCone" by (simp add: eCone\_def)
\\
}

\subsection*{Step 3. Identify the tangent plane containing \Code{f}}

Step 1 tells us to assume that \Code{f} is outside the cone. We can use the cone axioms to find a tangent plane containing \Code{f}. Being a tangent plane, it will necessarily contain the vertex, \Code{e}, as well. In addition, the axioms allow us to fix a point \Code{g} so that the line joining \Code{g} to the vertex is the line of intersection between the cone and the tangent plane. Notice that \Code{g} is distinct from both \Code{e} and \Code{f}, and together the three points define the tangent plane.

\Code{
~
\\  \indent have e\_is\_vertex: "e = vertex eCone" by (simp add: eCone\_def)
\\  \indent have cm\_is\_slope: "c m = slope eCone" by (simp add: eCone\_def)
\\  \indent hence outside: "outsideCone f eCone"
\\  \indent \indent by (metis (lifting) e\_is\_vertex cm\_is\_slope converse outsideCone.simps)
\\
\\  \indent have "outsideCone f eCone 
\\  \indent \indent $\longrightarrow$ ($\exists$x.(onCone x eCone $\land$ x $\noteq$ vertex eCone 
\\  \indent \indent \indent $\land$ inPlane f (tangentPlane x eCone)))"
\\  \indent by (rule AxParallelConesE)
\\
\\  \indent hence tplane\_exists: "$\exists$x.(onCone x eCone $\land$ x $\noteq$ vertex eCone 
\\  \indent \indent \indent $\land$ inPlane f (tangentPlane x eCone))" 
\\  \indent \indent by (smt outside)
\\  \indent then obtain g where g\_props: "(onCone g eCone $\land$ g $\noteq$ vertex eCone 
\\  \indent \indent \indent $\land$ inPlane f (tangentPlane g eCone))" 
\\  \indent \indent by auto
\\  \indent have g\_on\_eCone: "onCone g eCone" by (metis g\_props)
\\  \indent have g\_not\_vertex: "g $\noteq$ vertex eCone" by (metis g\_props)
\\
\\  \indent (* ... and more ... *)
\\
}

\subsection*{Step 4. Switch to \Code{k}'s viewpoint}
Because \Code{m} sees \Code{k} at the distinct points \Code{e} and \Code{f}, \Code{k} should also see himself at (his transformed versions of) those points, by \Code{AxEv}. But each observer considers himself to be stationary, so \Code{k} considers \Code{e} and \Code{f} to be distinct points on his time axis, by \Code{AxSelf}. If \Code{k}'s worldline also passed through \Code{g}, the points \Code{e}, \Code{f} and \Code{g} would be collinear in \Code{k}'s worldview, and hence also in \Code{m}'s, and we know this not to be the case because \Code{e} and \Code{g} are both in the tangent intersection line, while \Code{f} is outside the cone. Consequently, \Code{g} is not on \Code{k}'s time axis.

\Code{
~
\\  \indent def wvte $\equiv$ "wvt k m e"
\\  \indent def wvtf $\equiv$ "wvt k m f"
\\  \indent def wvtg $\equiv$ "wvt k m g"
\\
\\  \indent have "W k k wvte" by (metis wvte\_def AxWVT mke iobm iobk)
\\  \indent hence wvte\_onAxis: "onAxisT wvte" by (metis AxSelf iobk)
\\
\\  \indent have "W k k wvtf" by (metis wvtf\_def AxWVT mkf iobm iobk)
\\  \indent hence wvtf\_onAxis: "onAxisT wvtf" by (metis AxSelf iobk)
\\
\\  \indent have wvte\_inv: "e = wvt m k wvte" by (metis AxWVTSym iobk iobm wvte\_def)
\\  \indent have wvtf\_inv: "f = wvt m k wvtf" by (metis AxWVTSym iobk iobm wvtf\_def)
\\  \indent have wvtg\_inv: "g = wvt m k wvtg" by (metis AxWVTSym iobk iobm wvtg\_def)
\\
\\  \indent have e\_not\_g: "e $\noteq$ g" by (metis e\_is\_vertex g\_not\_vertex)
\\  \indent have f\_not\_g: "f $\noteq$ g" by (metis outside lemOutsideNotOnCone g\_on\_eCone)
\\
\\  \indent have wvt\_e\_not\_f: "wvte $\noteq$ wvtf" by (metis wvte\_inv wvtf\_inv enotf)
\\  \indent have wvt\_f\_not\_g: "wvtf $\noteq$ wvtg" by (metis wvtf\_inv wvtg\_inv f\_not\_g)
\\  \indent have wvt\_g\_not\_e: "wvtg $\noteq$ wvte" by (metis wvtg\_inv wvte\_inv e\_not\_g)
\\
\\  \indent have if\_g\_onAxis:  "onAxisT wvtg $\longrightarrow$ collinear wvte wvtg wvtf" 
\\  \indent \indent by (metis lemAxisIsLine wvte\_onAxis wvtf\_onAxis
\\  \indent \indent \indent \indent wvt\_e\_not\_f wvt\_f\_not\_g wvt\_g\_not\_e)
\\
\\  \indent have "collinear wvte wvtg wvtf $\longrightarrow$ collinear e g f"
\\  \indent \indent by (metis AxLines iobm iobk wvte\_inv wvtf\_inv wvtg\_inv)
\\  \indent hence "onAxisT wvtg $\longrightarrow$ collinear e g f" by (metis if\_g\_onAxis)
\\
\\  \indent hence wvtg\_offAxis: "$\lnot$ (onAxisT wvtg)" by (metis g\_not\_collinear)
\\ 
}

\subsection*{Step 5. Find a point \Code{z} with impossible properties}
We have seen that \Code{e} and \Code{f} define the time axis (from \Code{k}'s point of view), and \Code{g} lies off this axis. Consequently, because all lightcones are upright in special relativity, the line joining \Code{e} to \Code{g} has non-empty intersection with the \Code{k}-lightcone at \Code{f}. Call the point of intersection \Code{z}, and observe that the \Code{k}-lightcone at \Code{z} contains both \Code{e} and \Code{f}. [Notice, however, that determining the coordinates of the point \Code{z} typically involves the use of square roots, which is why we have assumed \Code{AxEuclidean}.] 

Having obtained \Code{z}, we will prove that its properties are contradictory. 

\Code{
~
\\  \indent have "$\forall$s.($\exists$p.( collinear wvte wvtg p 
\\  \indent \indent \indent \indent $\land$ (space2 p wvtf = (s*s)*time2 p wvtf)))"
\\  \indent \indent by (metis lemSlopedLineInVerticalPlane 
\\  \indent \indent \indent wvte\_onAxis wvtf\_onAxis wvtg\_offAxis wvt\_e\_not\_f)
\\  \indent hence exists\_wvtz: "$\exists$p.( collinear wvte wvtg p 
\\  \indent \indent \indent \indent $\land$ (space2 p wvtf = (c k * c k)*time2 p wvtf))"
\\  \indent \indent by metis
\\  \indent then obtain wvtz where 
\\  \indent \indent wvtz\_props: "collinear wvte wvtg wvtz 
\\  \indent \indent \indent $\land$ (space2 wvtz wvtf = (c k * c k)*time2 wvtz wvtf)" by auto
\\  \indent hence wvtf\_speed: "space2 wvtz wvtf = (c k * c k)*time2 wvtz wvtf" 
\\  \indent \indent by metis
\\
\\  \indent def z $\equiv$ "wvt m k wvtz"
\\
}

We know that \Code{f} is on \Code{k}'s lightcone at \Code{z}, and that lightcones are mapped to lightcones under worldview transformations. We can therefore switch to \Code{m}'s viewpoint, and at the same time deduce that \Code{z} is on the lightcone at \Code{f}.

\Code{
~
\\  \indent (* f is on the lightcone at z *)
\\  \indent def zCone $\equiv$ "lightcone m z"
\\  \indent have z\_is\_vertex: "z = vertex zCone" by (simp add: zCone\_def)
\\  \indent have cm\_is\_zSlope: "c m = slope zCone" by (simp add: zCone\_def)
\\
\\  \indent have f\_on\_zCone: "onCone f zCone" 
\\  \indent \indent by (metis wvtf\_inv wvtf\_on\_wvtzCone zCone\_def)
\\
\\  \indent (* whence z is on the lightcone at f *)
\\  \indent hence "space2 (vertex zCone) f 
\\  \indent \indent \indent = (slope zCone * slope zCone)*time2 (vertex zCone) f"
\\  \indent \indent by (simp add: zCone\_def)
\\  \indent hence "space2 z f = (c m * c m)*time2 z f" 
\\  \indent \indent by (metis z\_is\_vertex cm\_is\_zSlope)
\\  \indent hence fz\_speed: "space2 f z = (c m * c m)*time2 f z" 
\\  \indent \indent by (metis lemSpace2Sym lemTime2Sym)
\\
\\  \indent def fCone $\equiv$ "lightcone m f"
\\  \indent have f\_is\_fVertex: "f = vertex fCone" by (simp add: fCone\_def)
\\  \indent have cm\_is\_fSlope: "c m = slope fCone" by (simp add: fCone\_def)
\\  \indent hence "space2 (vertex fCone) z 
\\  \indent \indent \indent = ((slope fCone) *(slope fCone))*time2 (vertex fCone) z" 
\\  \indent \indent by (metis fz\_speed f\_is\_fVertex cm\_is\_fSlope)
\\  \indent hence z\_on\_fCone: "onCone z fCone" by (metis onCone.simps)
\\
}

Similarly, we can show that \Code{z} is on the lightcone at \Code{e}. However, the cones at \Code{e} and \Code{f} share the same tangent plane (because \Code{f} lies in that plane), whence the intersection lines at \Code{e} and \Code{f} are parallel (this is part of what it means to be a tangent plane, as expressed in the cone axioms). It follows that we have two distinct lines that intersect in a common point, \Code{z}, despite being parallel.

This provides the required contradiction.

\section{Discussion}
\label{sec:discussion}

In practice, the most time-consuming part of this proof involved describing the geometric properties of spacetime -- for example, deciding the best way to represent lines and planes, what it means for points to be collinear or coplanar, or what it means for two lines to be parallel. This suggests that Isabelle/HOL should provide an excellent vehicle for constructing future proofs relating to the more complex versions of relativity theory, because all standard models of general relativity are locally special relativistic. Consequently, we expect that work already invested in the construction of \Code{SpaceTime.thy} (itself built on top of existing Isabelle/HOL libraries) will largely be re-usable.

There remains, of course, a great deal more to be done. In addition to completing the proofs of other standard features of special relativity (for example, time dilation), we need to extend our work to both accelerating observers and their associated theorems (for example, the ``twin paradox''), and observers in a gravitational field. Only then will we be in a position to model what it means for a spacetime to exhibit the Malament-Hogarth timing structures relevant to existing suggestions for cosmological (hyper)computation. We also plan to continue the investigation into the physical realisticity of computing with Malament-Hogarth spacetimes started in \cite{ND06,AN06}, not necessarily sticking with Kerr spacetime (cf. \cite{Man10}).

Finally, we would like to know to what extent the work developed here can be extended to encompass other physical systems -- for example quantum mechanics -- and whether new proof techniques or capabilities would be useful in that effort. For example, in the proof above it was necessary for us to determine the existence of a point \Code{z} with certain coordinates. Although it was straightforward to compute those coordinates by hand, it would be convenient to have a system built into Isabelle/HOL that could do the construction on our behalf, or at least tell us whether a suitable point \Code{z} exists.

\section*{Acknowledgements}
This research is supported under the Royal Society International Exchanges Scheme (ref. IE110369). N\'emeti's research was supported by OTKA grant No 81188. This work was partially undertaken whilst Stannett was a visiting fellow at the Isaac Newton Institute for the Mathematical Sciences in the programme \textit{Semantics \& Syntax: A Legacy of Alan Turing}.

\bibliographystyle{alpha}
\bibliography{verispec}

\end{document}